\documentclass[fleqn,usenatbib]{mnras}
\usepackage{mathptmx}
\usepackage[varvw]{newtxmath}  
\usepackage[T1]{fontenc}
\usepackage{ae,aecompl}
\usepackage{times}
\usepackage{amsmath}
\usepackage{upgreek}
\usepackage{xcolor}
\usepackage{graphicx}
\usepackage{pdflscape}
\usepackage{multicol} 
\usepackage{amsmath,bm}
\usepackage{caption}

\begin{document}

\title{Evolution of molecular clouds on galaxy-cloud scale revealed by gravitational network analysis : High-mass clouds may deplete nearby
gas via accretion or merging}
\author[J. W. Zhou]{
J. W. Zhou \thanks{E-mail: jwzhou@mpifr-bonn.mpg.de}$^{1}$
Guang-Xing Li \thanks{E-mail: ligx.ngc7293@gmail.com}$^{2}$
\\
$^{1}$Max-Planck-Institut f\"{u}r Radioastronomie, Auf dem H\"{u}gel 69, 53121 Bonn, Germany\\
$^{2}$South-Western Institute For Astronomy Research, Yunnan University, Kunming 650600 Yunnan, P.R., China
}

\date{Accepted XXX. Received YYY; in original form ZZZ}

\pubyear{2024}
\maketitle

\begin{abstract}
Observations show that molecular gas in spiral galaxies is organized into a network of interconnected systems through the gravitational coupling of multi-scale hub-filament structures. Building on this picture, we model molecular gas in the galaxy NGC 628 as a gravitational network, where molecular clouds are represented as nodes. 
Through analyzing this network,
we can characterize both the gravitational interactions and the physical properties of the clouds using geometry-based network metrics. 
A strong correlation is observed between the geometric and physical properties of the nodes (clouds). High-mass clouds tend to exhibit less clustering and greater average separations, suggesting that they generally have fewer neighbors. During their formation and evolution, high-mass clouds may deplete nearby gas via accretion or merging, leading to more isolated characteristics within the network. This aligns with observations showing a decrease in the virial ratio of molecular clouds as their mass increases. For clouds at different evolutionary stages, less evolved clouds with lower mass are typically found in tighter gravitational subnetworks, with closer proximity to neighboring clouds. As a result, they are more prone to accretion or merging during evolution. 

\end{abstract}


\section{Introduction} 

Giant molecular clouds (GMCs) are widely acknowledged as the primary reservoirs of gas that drive star formation, serving as the cradles for nearly all stars. Understanding the properties of GMCs is therefore essential for unraveling the intricate link between gas and star formation in galaxies \citep{Kennicutt2012-50,Schinnerer2024-62}. Observations show that the physical characteristics of GMCs systematically change depending on their location within a galaxy, underscoring the significant interaction between these clouds and their surrounding environment, which influences their formation, structure, and evolution \citep{Dib2012-758, Hughes2013-779,Colombo2014-784,Miville2017-834,Sun2022-164,Kim2022-516,Pessa2022-663,Zhou2024-534}.

Kinematic studies reveal the existence of multi-scale, hierarchical hub-filament structures within molecular clouds and spiral galaxies \citep{Peretto2013,Lu2018,Henshaw2020-4,Zhou2022-514,Zhou2023-676,Zhou2024PASA}. Their findings particularly highlight that intensity peaks, functioning as hubs, are linked to converging velocities, implying that the surrounding gas is flowing towards these dense regions. Additionally, the observed variations in the velocity gradient across different scales point to a gradual and consistent increase from larger to smaller scales, which is indicative of gravitational collapse. 
Especially,
the velocity gradients on larger scales require a higher central mass to be maintained. Higher masses correspond to larger scales, suggesting that large-scale inflow is driven by these larger structures. This could be a consequence of the gravitational clustering of smaller-scale structures, implying hierarchical gas structures and gas inflow from larger to smaller scales.
Based on the kinematic evidence, \citet{Zhou2024-534} directly decomposed the molecular gas in the spiral galaxy NGC 628 into multi-scale hub-filament structures using the CO (2-1) line map. The molecular clouds identified as potential hubs were categorized into three groups: leaf-HFs-A, leaf-HFs-B, and leaf-HFs-C. For leaf-HFs-C, the contrast in density between the hubs and the surrounding diffuse gas is not pronounced. Both leaf-HFs-A and leaf-HFs-B show distinct central hubs, but leaf-HFs-A exhibit the most well-defined hub-filament structure. Leaf-HFs-A have the highest density contrast, the largest mass, and the lowest virial ratio. 
It is possible that there is an evolutionary sequence from leaf-HFs-C to leaf-HFs-A. Currently, leaf-HFs-C do not show a clear gravitational collapse process that would lead to a high density contrast.
These observations suggest that molecular gas in spiral galaxies is structured into a network of interconnected systems through the gravitational coupling of multi-scale hub-filament structures. Molecular clouds act as nodes within these hub-filament networks, serving as local gravitational centers and primary sites for star formation. 
In this work, we employ network-based analytical methods to gain further insights into the formation and evolution of molecular clouds in the spiral galaxy NGC 628.

\section{Data}
In \citet{Zhou2024-534},
we used the combined 12m+7m+TP PHANGS-ALMA CO (2$-$1) data cube of the spiral galaxy NGC 628 to investigate gas kinematics and dynamics and identify hub-filament structures. The CO data has a spectral resolution of 2.5 km s$^{-1}$ and an angular resolution $\sim$1.1$''$, corresponding
to a linear resolution $\sim$50 pc at the distance 9.8 Mpc \citep{Leroy2021-257,Anand2021-501}. All the data are available on the PHANGS team website \footnote{\url{https://sites.google.com/view/phangs/home}}. The analysis in this work is based on the cloud catalog constructed in \citet{Zhou2024-534}.

\section{Results and Discussion}

\subsection{Gravitational network}\label{gn}

\begin{figure*}
\centering
\includegraphics[width=0.95\textwidth]{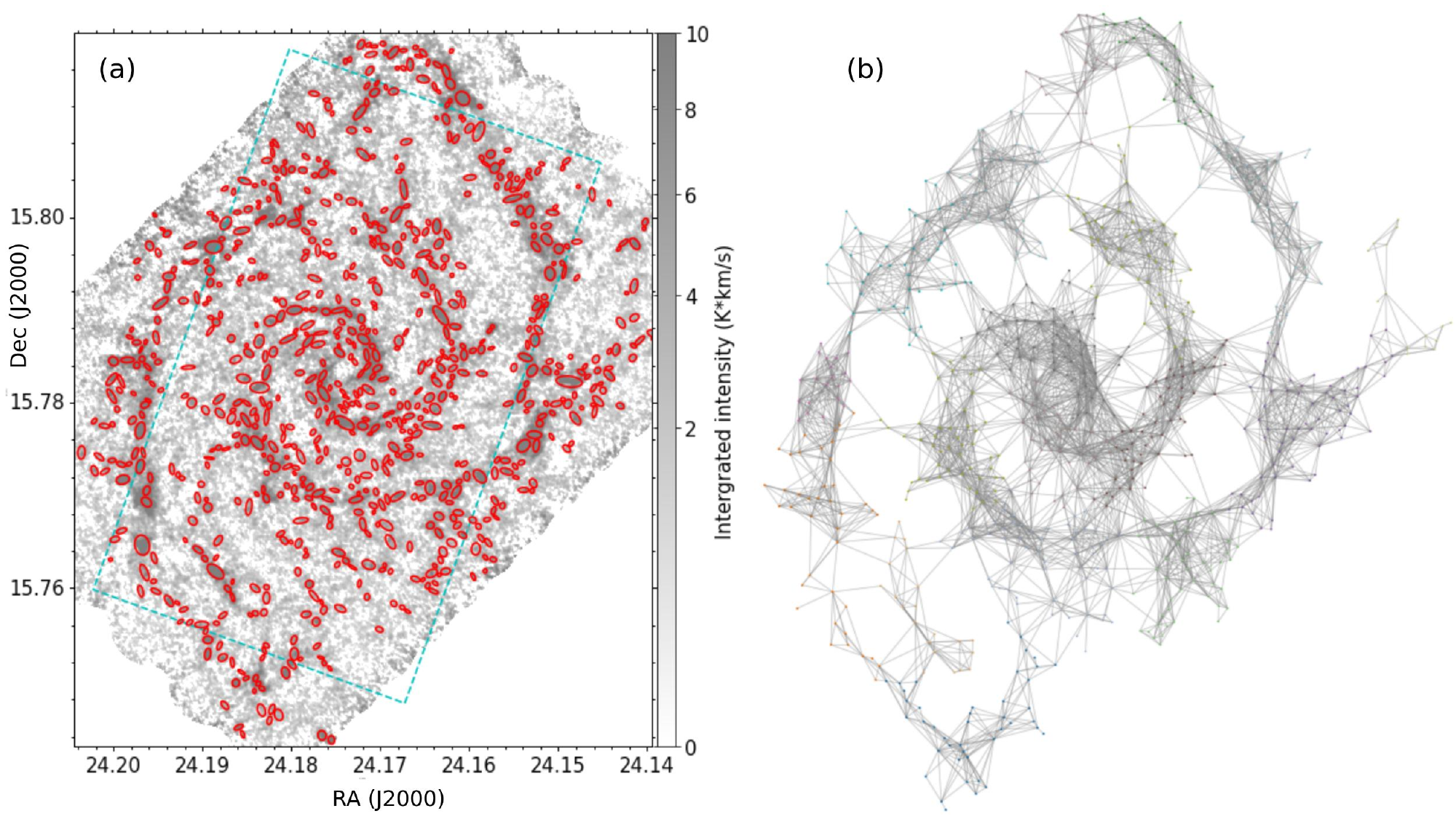}
\caption{(a) Background is the broad-masked integrated intensity map of CO (2$-$1) emission. Red ellipses represent leaf structures identified by the dendrogram algorithm in \citet{Zhou2024-534}. Cyan box marks the field-of-views of the JWST 21 $\mu$m observation; (b) The gravitational network constructed in Sec.\ref{gn}, with molecular clouds as nodes.}
\label{network}
\end{figure*}

Considering the picture of multi-scale/hierarchical hub-filament structures in spiral galaxies revealed by observations, we can construct a gravitational network with molecular clouds as nodes all over the galaxy NGC 628.
The elements of a network include nodes, edge weights and connection rules. The edge weights are defined by the gravitational force between nodes, with greater weights indicating stronger gravitational interactions. The connection rules stipulate that nodes separated by a distance greater than 1 kpc are not connected, as this work focuses only on locally coupled structures. The distance is calculated in a 2D plane using Euclidean distance according to the central coordinates of nodes. 
We tested different values of the separation threshold and found that a threshold below 1 kpc (e.g., 500 pc) mainly results in discrete local networks, preventing the formation of the overall network shown in Fig.\ref{network}. A larger threshold (e.g., 1500 pc) leads to more connections, which eliminate locally coupled structures (local networks) but do not affect the correlations between node geometric parameters and node mass presented in the next section.
Furthermore, a gravitational threshold is defined as the median gravitational force calculated for all node pairs with separations less than 1 kpc.
Nodes are connected only if their gravitational force exceeds this threshold to outstand the main gravitational interactions between nodes.
Fig.\ref{network} presents the constructed network, revealing some localized cloud clusters.

\subsection{Accretion or merger during the formation and evolution of molecular clouds}

\begin{figure*}
\centering
\includegraphics[width=0.9\textwidth]{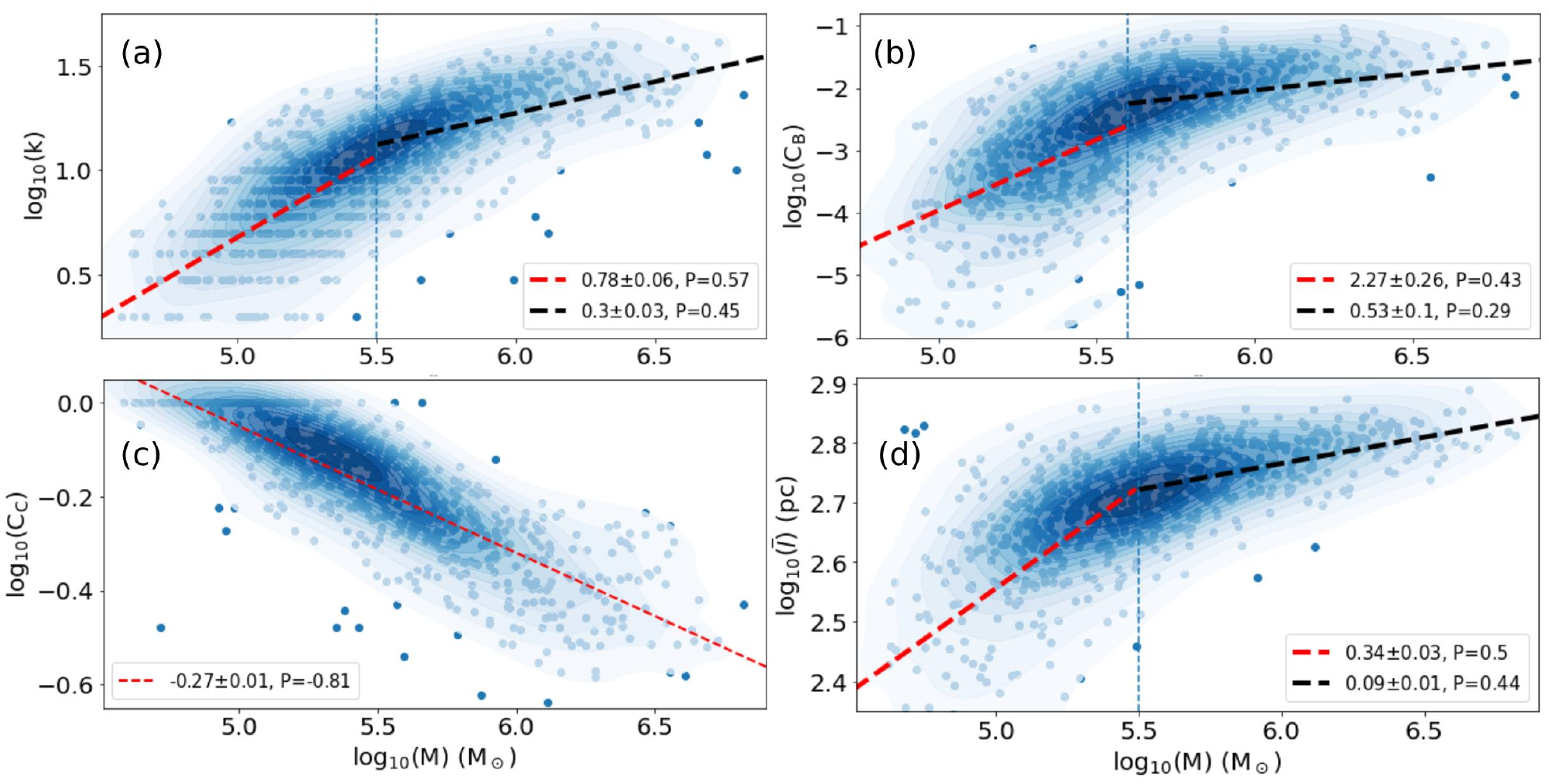}
\caption{Correlation between the geometric parameters of nodes and node mass. (a) The degree ($k$); (b) The betweenness centrality ($C_{B}$); (c) The clustering coefficient ($C_{C}$); (d) The average physical path length ($\overline{l}$) between each node and its neighbors. The background contour in each panel shows the probability density (KDE) of the scatter points. Vertical dashed lines mark the inflection points of the segmented fit.}
\label{para}
\end{figure*}

\begin{figure*}
\centering
\includegraphics[width=1\textwidth]{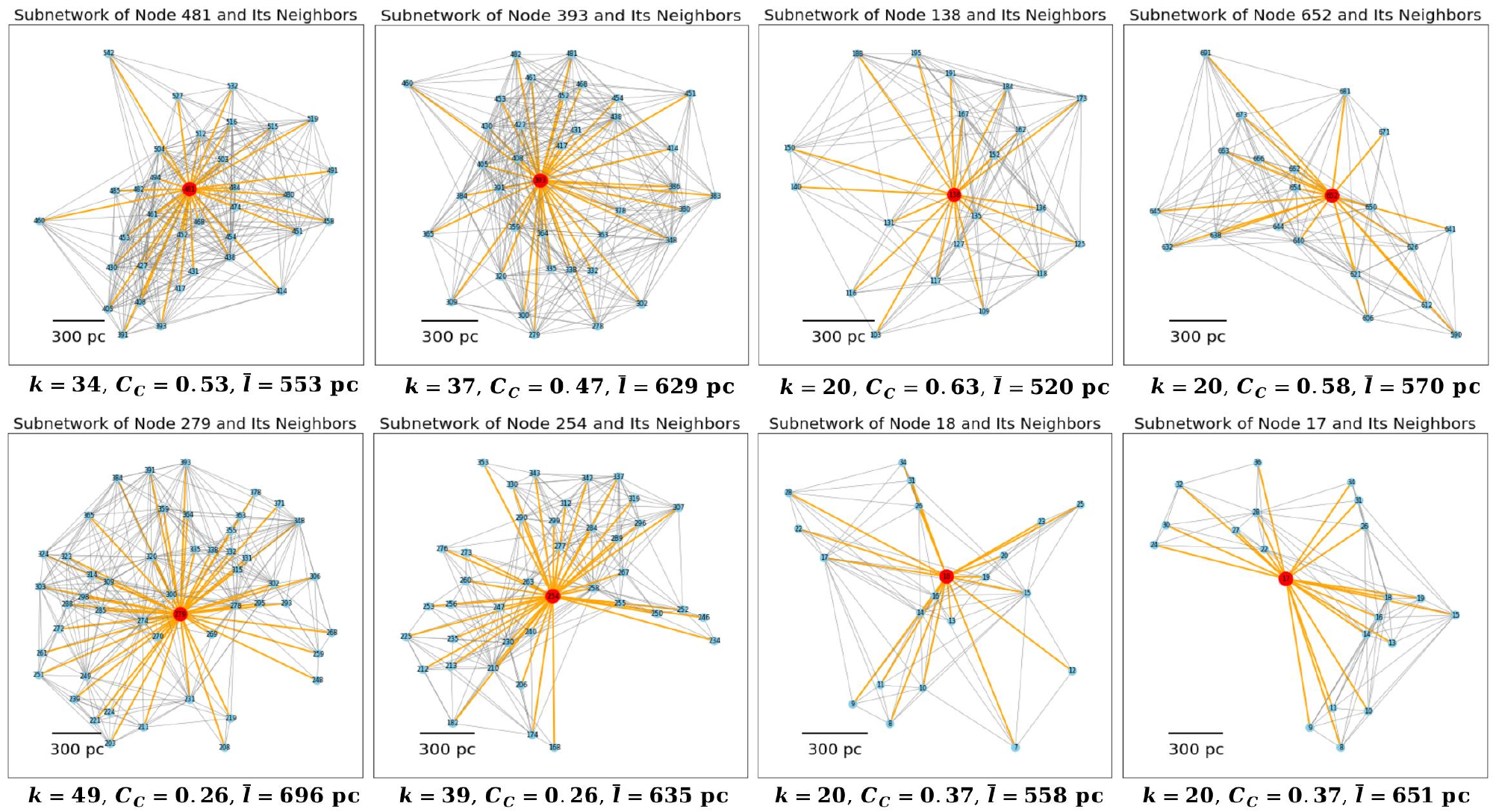}
\caption{Some subnetworks and their geometric properties. Red dot and blue dots represent the node and its neighbors, respectively. $k$, $C_{C}$ and $\overline{l}$ are the degree, the clustering coefficient and the average physical path length between each node and its neighbors, respectively.}
\label{sub}
\end{figure*}

\begin{figure}
\centering
\includegraphics[width=0.45\textwidth]{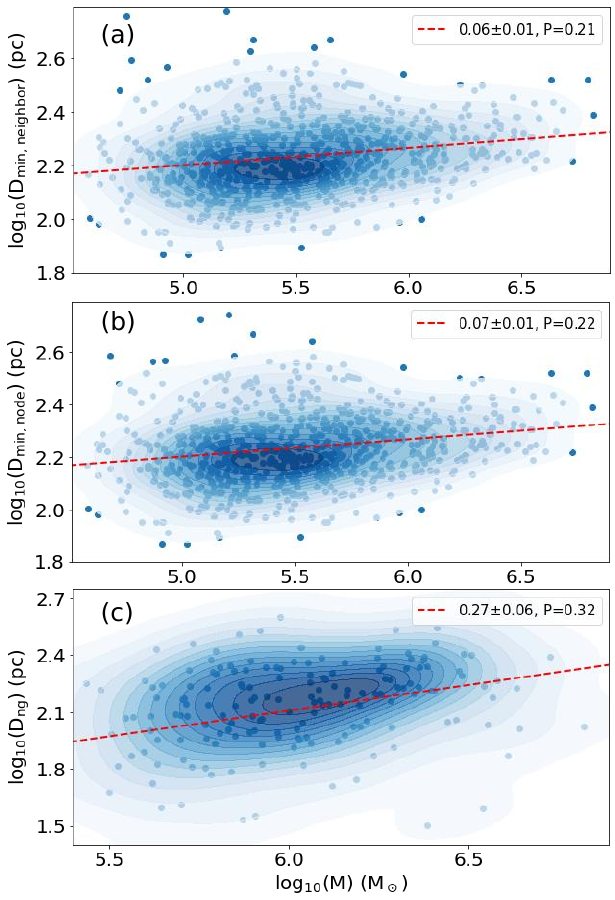}
\caption{(a) Correlation between node mass and the distance to its nearest neighbor ($D_{min,neighbor}$); (b) Correlation between node mass and the distance to its nearest node ($D_{min,node}$); (c) Correlation between node mass and the displacement of the node from the geometric center of the subnetwork comprising the node and its neighbors ($D_{ng}$). To ensure more precise calculations of the subnetwork's geometric center, we only consider nodes with a degree of $\geq$20. The background contour in each panel shows the probability density (KDE) of the scatter points.}
\label{dis}
\end{figure}

\begin{figure}
\centering
\includegraphics[width=0.45\textwidth]{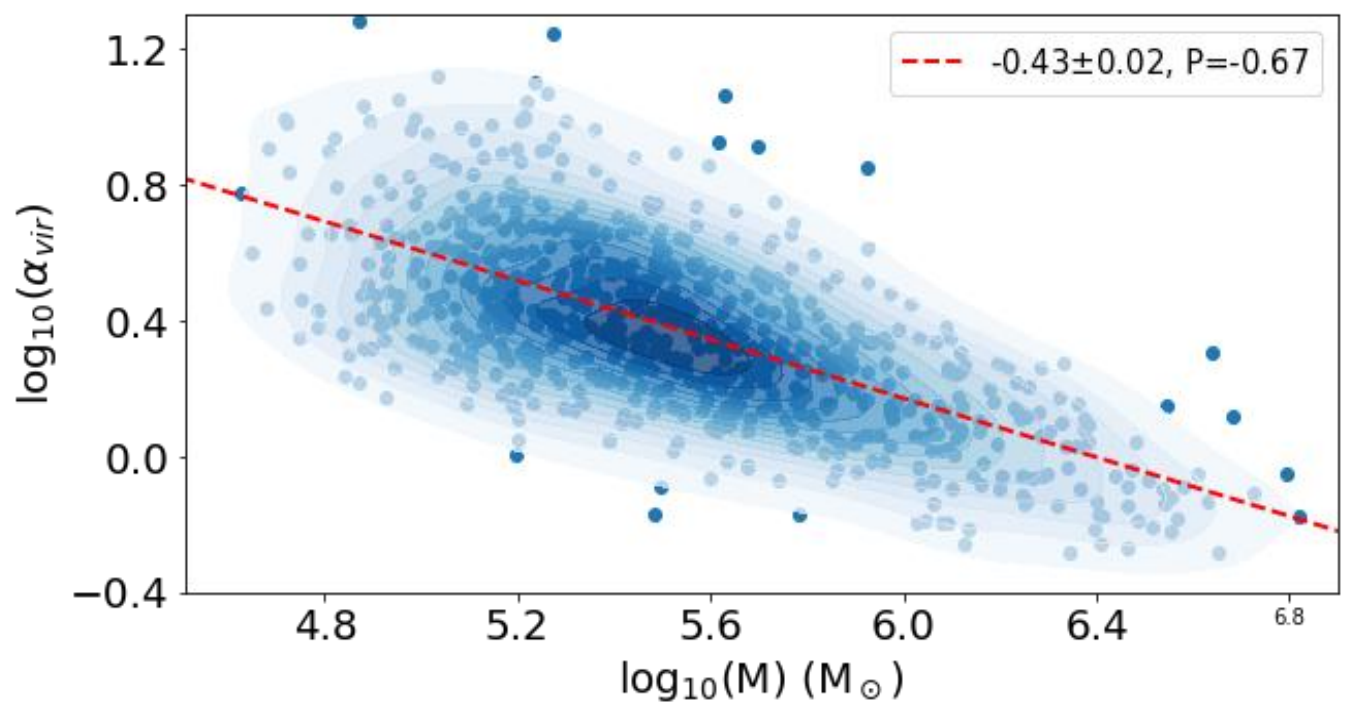}
\caption{Correlation between the virial ratio and mass of molecular clouds. The background contour shows the probability density (KDE) of the scatter points.}
\label{virial}
\end{figure}

\begin{figure}
\centering
\includegraphics[width=0.45\textwidth]{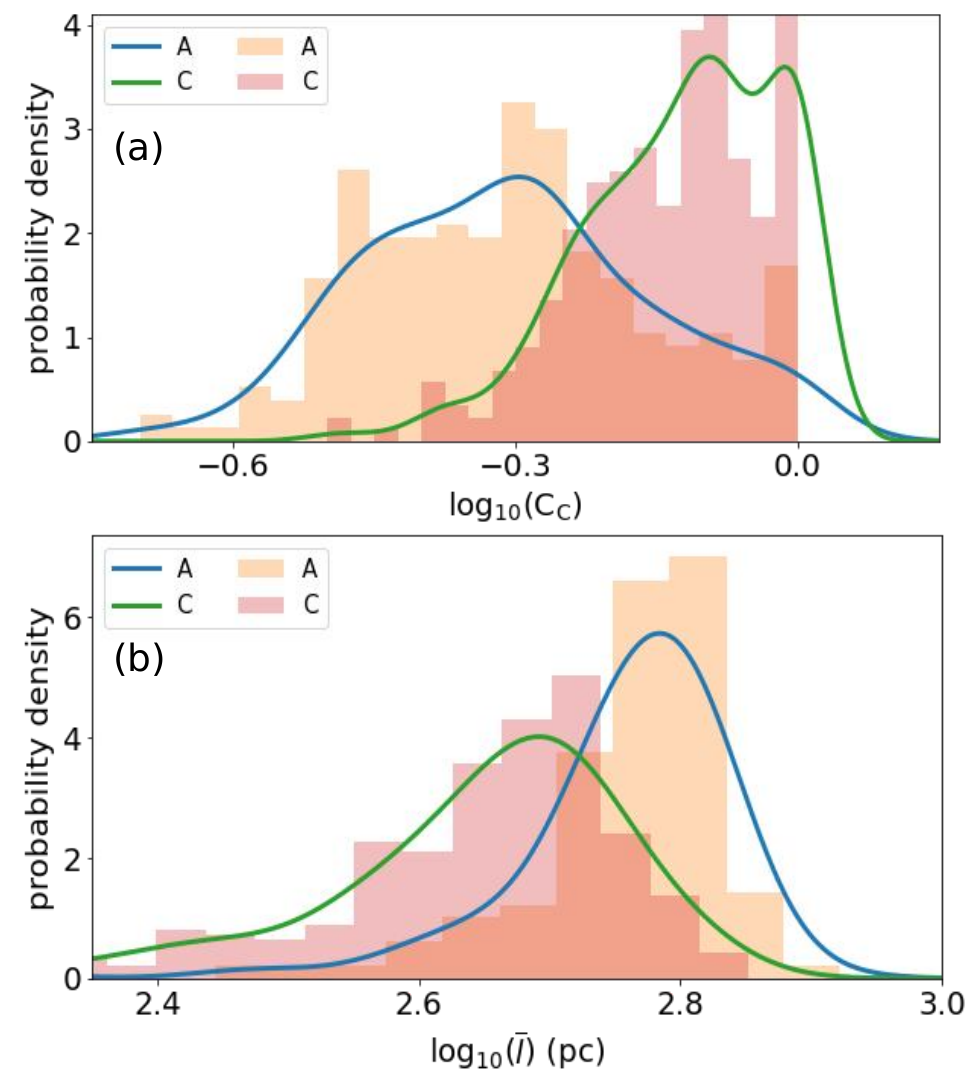}
\caption{Comparison of the geometric parameters of A-type and C-type clouds. (a) The clustering coefficient ($C_{C}$); (b) The average physical path length  between each node and its neighbors ($\overline{l}$).}
\label{type}
\end{figure}

The physical properties of molecular clouds in NGC 628 have been calculated and discussed in \citet{Zhou2024-534}. Now, with each cloud being considered a node, we can calculate the geometric parameters of each node and explore the correlation between geometric and physical properties of molecular clouds. In other words, since we have visualized the gravitational interactions between molecular clouds using the gravitational network, we can now characterize the gravitational interactions between the clouds or the physical properties of the clouds themselves using geometry-based network metrics.

In a network, the degree of a node ($k_{i}$) represents the number of other nodes directly connected to this node. It is a fundamental attribute for measuring the importance of a node. Particularly in a gravitational network, nodes with high degrees may represent massive objects that exert significant gravitational influence on their surroundings. Fig.\ref{para}(a)
shows a strong positive correlation between the mass of the nodes and their degree.

The betweenness centrality of a node ($C_{B,i}$, see the definition of equation.\ref{between}) measures the extent to which this node acts as an intermediary within the network. Nodes with high betweenness centrality have greater influence in the network. Massive clouds, serving as local hubs and gravitational centers, naturally exhibit greater influence, leading to higher values of $C_{B}$, as shown in Fig.\ref{para}(b).

The clustering coefficient of a node ($C_{C,i}$, see the definition of equation.\ref{cc}) is an important metric for measuring the degree of connectivity among its neighbors. It reflects the tightness of the local network structure around a node. In a gravitational network, a high clustering coefficient indicates that the neighbors of a node have stronger gravitational relationships with one another, forming a highly interconnected subnetwork. Surprisingly, as shown in Fig.\ref{para}(c), high-mass molecular clouds exhibit a lower clustering coefficient. 
We also calculate the average physical path length ($\overline{l_{i}}$) between each node and its neighbors (the average length of all orange lines for each subnetwork in Fig.\ref{sub}). In Fig.\ref{para}(d), $\overline{l}$ increases with the node's mass, which means an anti-correlation between $\overline{l}$ and $C_{C}$. This could be due to high-mass nodes connecting to more distant nodes, which increases the average distance to their neighbors. Alternatively, it might be that there are simply fewer nodes near high-mass nodes, which also accounts for the lower $C_{C}$ observed in these high-mass nodes.

Fig.\ref{dis}(a) shows the relationship between node mass and the distance to its nearest neighbor ($D_{min}$), demonstrating a positive correlation between node mass and its distance from the nearest neighbor. 
If we relax the requirement and only consider the shortest distance between two nodes, without requiring them to be neighbors, we observe a similar correlation between node mass and distance (Fig.\ref{dis}(b)).
In Fig.\ref{dis}(c), for subnetworks formed by a node and its neighbors (degree $\geq$ 20), nodes with greater mass and longer average path lengths tend to show a more significant displacement from the geometric center of the subnetwork ($D_{ng}$). Fig.\ref{sub} displays some subnetworks and their geometric properties.

The above results suggest that high-mass clouds, during their formation and evolution, deplete nearby gas through accretion or merger, resulting in more isolated network characteristics. 
Molecular clouds are inherently irregular gas structures, so describing them as being merged or accreted essentially conveys a similar idea. 
Initially, molecular clouds may exhibit a relatively uniform distribution. Over time, however, molecular gas accumulates towards the gravitational centers, and the high-mass clouds that form eventually become decoupled from their surrounding environment, as these clouds experience stronger self-gravity. As shown in Fig.\ref{virial}, the virial ratio of molecular clouds decreases as the molecular cloud mass increases.


To further verify the above scenario, now we focus on two different types of molecular clouds classified in \citet{Zhou2024-534}, i.e. leaf-HFs-A and leaf-HFs-C (A-type and C-type). According to the evolutionary sequence proposed in \citet{Zhou2024-534}, A-type clouds are more evolved than C-type clouds. As shown in Fig.\ref{type}, C-type clouds as nodes have higher $C_{\rm C}$ and smaller $\overline{l}$. This indicates that, compared to A-type clouds, C-type clouds are generally located in tighter gravitational subnetworks, with shorter distances to other molecular clouds, which makes them more prone to accretion or merging as they evolve.
As shown in \citet{Zhou2024-534}, the more evolved A-type clouds have significantly larger mass and lower virial ratio than C-type clouds. In Fig.\ref{para}(d), for the relation $M \propto (\overline{l})^{\alpha}$, low-mass clouds exhibit a significantly steeper slope ($\alpha \approx 0.34$) compared to high-mass clouds ($\alpha \approx 0.1$), suggesting that low-mass clouds are more susceptible to accretion or merging during evolution, consistent with the above results. 


\section{Conclusion}

Observations reveal that molecular gas in spiral galaxies forms a network of interconnected systems through the gravitational coupling of multi-scale hub-filament structures. Within these networks, molecular clouds serve as nodes, acting as local gravitational centers and primary sites for star formation. Based on this physical picture, we construct a gravitational network for the galaxy NGC 628, where molecular clouds are treated as nodes.
By visualizing the gravitational interactions between molecular clouds in this network, we can characterize both the gravitational interactions and the physical properties of the clouds using geometry-based network metrics. In the gravitational network, there is a strong correlation between the geometric and physical properties of the nodes (clouds). Metrics such as node degree ($k$), betweenness centrality ($C_B$), clustering coefficient ($C_C$), and average physical path length ($\overline{l}$) show strong correlations with node mass. 
High-mass molecular clouds exhibit lower $C_C$ and longer $\overline{l}$, indicating that high-mass nodes tend to have fewer neighbors. Additionally, there is a positive correlation between node mass and the distance to the nearest neighbor ($D_{\text{min}}$). For subnetworks consisting of a node and its neighbors (node degree $\geq 20$), nodes with greater mass and longer $\overline{l}$ are more significantly displaced from the geometric center of the subnetwork ($D_{\text{ng}}$).

These findings suggest that high-mass clouds, during their formation and evolution, deplete nearby gas through accretion or merger, resulting in more isolated network characteristics.
This is consistent with the observed decrease in the virial ratio of molecular clouds as their mass increases.
For molecular clouds at different evolutionary stages classified in \citet{Zhou2024-534} (i.e. A-type and C-type), a comparison reveals that C-type clouds, which are less evolved, have higher $C_C$ and shorter $\overline{l}$ compared to A-type clouds. This indicates that C-type clouds are generally located in tighter gravitational subnetworks, with closer proximity to other molecular clouds. Consequently, they are more prone to accretion or merging during evolution. For the relation $M \propto (\overline{l})^{\alpha}$, the slope for low-mass clouds ($\alpha \approx 0.34$) is significantly steeper than that for high-mass clouds ($\alpha \approx 0.1$). This also indicates that low-mass clouds are more susceptible to accretion or merging as they evolve, as the more evolved A-type clouds possess significantly greater mass and a lower virial ratio compared to C-type clouds.

\section{Data availability}

All the data used in this work are available on the PHANGS team website.
\footnote{\url{https://sites.google.com/view/phangs/home}}.


\bibliography{ref}
\bibliographystyle{aasjournal}

\begin{appendix}
\twocolumn

\section{Definition of geometric parameters}

Betweenness centrality measures the extent to which a node acts as an intermediary or "broker" within a network. It quantifies how often a node lies on the shortest path between two other nodes, reflecting its role in connecting different parts of the network.
Formally, the betweenness centrality $C_{B,i}$ of a node $i$ is defined as:
\begin{equation}
    C_{B,i} = \sum_{s \neq i \neq t} \frac{\sigma_{\rm st}(i)}{\sigma_{\rm st}}
\label{between}
\end{equation}
where $\sigma_{\rm st}$ is the total number of shortest paths between nodes $s$ and $t$.
$\sigma_{\rm st}(i)$ is the number of shortest paths between nodes $s$ and $t$ that pass through node $i$.
The sum is taken over all pairs of nodes $s$ and $t$ that are not equal to $i$. A node with high betweenness centrality plays a crucial role in connecting different parts of the network, acting as a bridge between otherwise disconnected nodes or communities. Nodes with high betweenness centrality often have significant influence because they control the flow of information or resources across the network.

For a node $i$, the clustering coefficient is defined as the ratio of the actual number of edges between its neighbors to the total possible number of edges between them. It quantifies the tightness of the local network around the node. Specifically, the clustering coefficient $C_{C,i}$ of node $i$ can be calculated using the following formula:
\begin{equation}
    C_{C,i} = \frac{2e_{\rm i}}{k_{\rm i}(k_{\rm i} - 1)},
\label{cc}
\end{equation}
where $e_{\rm i}$ is the actual number of edges between the neighbors of node $i$ (i.e., the number of edges that exist between the neighbors). $k_{\rm i}$ is the degree of node $i$, i.e., the number of neighbors of node $i$. If the clustering coefficient $C_{C,i}$ is close to 1, it indicates that the neighbors of node $i$ are nearly all connected to each other, meaning the local network is very tight. If $C_{C,i}$ is close to 0, it indicates that there are almost no connections between the neighbors of node $i$, meaning the local network is sparse.

\end{appendix}

\clearpage
\noindent
\end{document}